# We.I.Pe: Web Identification of People using e-mail ID

K.S. Kuppusamy
Department of Computer Science, School of Engineering & Technology
Pondicherry University
Pondicherry, India

G.Aghila
Department of Computer Science, School of Engineering & Technology
Pondicherry University
Pondicherry, India

*Abstract*— **With the phenomenal growth of content in the World Wide Web, the diversity of user supplied queries have become vivid. Searching for people on the web has become an important type of search activity in the web search engines. This paper proposes a model named "We.I.Pe" to identify people on the World Wide Web using e-mail Id as the primary input. The approach followed in this research work provides the collected information, based on the user supplied e-mail id, in an easier to navigate manner. The grouping of collected information based on various sources makes the result visualization process more effective. The proposed model is validated by a prototype implementation. Experiments conducted on the prototype implementation provide encouraging results.**

*Keywords-information retrieval; web search engines; people search; e-mail id based search search engine results; personalization;*

I. INTRODUCTION

The World Wide Web has become the largest repository of information known to the mankind. Because of the quantity and diversity of nature of information available in the World Wide Web, it is been used as an easier and effective reference source for the people of all works of life. The web search engines make the process of accessing the information on the World Wide Web effective. A web search oriented towards location information regarding people is known as "people search". Searching for people has become one of the important search activities in the major web search engines. The research work [1] suggests that the people search, in certain cases of search engine logs, occupies up to 10% of the overall search activity.

Searching for people on the web has got certain unique requirements than searching an ordinary query in the search engines. This paper proposes a model to search for people on the web using their e-mail id as the primary factor.

The major objectives of this research work are as listed below:

- Proposing a model to search for people on the world wide which receives e-mail id as input and persons profile and related information as output.

- Providing an intuitive visualization by garnering information gathered using the people search.

The remainder of this paper is organized as follows: In section 2 the related works carried out in this domain which has formed the motivation for this work are listed out. Section 3 is about the proposed model and the algorithm. Section 4 focuses on prototype implementation carried out on the proposed model and the experimentation results. Section 5 charts out the conclusions and future directions for this research work.

II. MOTIVATIONS

The people search on the web is one of the active research topics in the web information retrieval domain. Though the people search can be carried out in the generic search engines itself, their effectiveness is limited due to the "one-size fits all" policy. There exist few initiatives from popular search engines like Yahoo to provide specialized interface to search for people.

The Yahoo People Search [2] enable to search people on the web based on their first name, last name etc. It also provides the telephone number based search. Most of these facilities are targeted towards users in United

States. The dedicated people search engines like Pipl, Intelius [3][4] focuses on providing detailed profiles based on various inputs.

A test bed for people search strategies on World Wide Web is explored in [5]. Another important issue with people search is the disambiguation of person names. A study based on "person cluster hypothesis" is explored in [6]. A robust disambiguation technique for people search is explored in [7]. Named entity based approaches are discussed in [8], [9].

Though the web holds information on people, social networks on the web consists of huge amount of information on people, their linkages etc. These social networks can be treated like a directed graph. An improved people-search technique for directed social network graphs is explored in [10].

The approach that has been followed in our research is to harness the power of these huge warehouses of information i.e. social networks to locate information about the person whose email id is provided. In addition to these social networks the searches are conducted on general web resources as well.

III. THE MODEL

The block diagram for the proposed We.I.Pe model is depicted in Figure.1. It receives the e-mail id of the person whose information is going to be located as input. Hence the information is gathered from both social networks and other resources, corresponding managers and parsers are there in the model. The result visualizer has components to display both summary information and blog profile information.

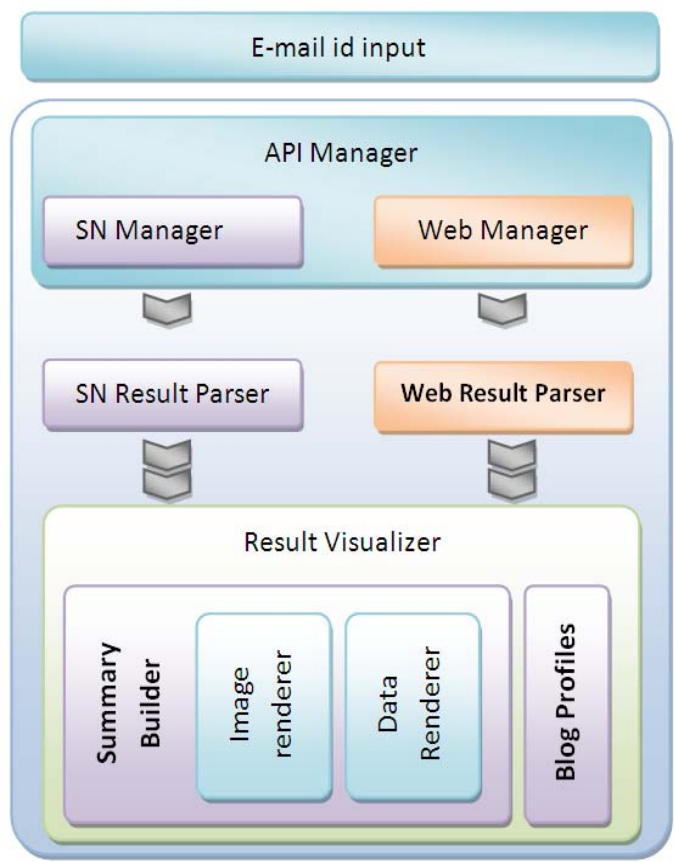

Figure 1. The We.I.Pe Model

*A. The mathematical Model*

The mathematical representation of the proposed We.I.Pe model is explored in this section. The input to the We.I.Pe model is the e-mail id for which information needs to be retrieved. The advantage with having this e-mail id as input is that any given e-mail id in the world is unique, i.e. no two persons can have the same e-mail id. This unique nature of e-mail id makes the results retrieved to be unique to the particular person having that e-mail id.

The e-mail id given as input can be denoted as $\delta$. After the submission of the e-mail id by the user, it is forwarded to the API (Application Programming Interface) manager $\Omega$. API manager is responsible for establishing connection to the information resources. The API manager has got two components as indicated in (1).

$$\Omega = \{\lambda, \pi\} \quad (1)$$

In (1), $\lambda$ represents the Social Network manager and $\pi$ represents the Web manager. The Social Network Manager $\lambda$ is responsible for establishing connection to the social networks and fetching the information from it. Similarly the Web Manager $\pi$ is responsible for locating content from World Wide Web.

$$\lambda = \left\{\frac{\Gamma}{\delta}\right\}^{\varepsilon} \quad (2)$$

In (2) $\Gamma$ denotes the Social Networking domain / resources.

$$\pi = \left\{\frac{\Pi}{\delta}\right\}^{\varepsilon} \quad (3)$$

In (3) $\Pi$ denotes the resources of World Wide Web excluding the contents of Social Networks. In both (2) and (3) the symbol $\varepsilon$ represents the threshold limit. This threshold limit indicates the top "n" results which need to be considered for the evaluation purpose of the model. This threshold $\varepsilon$ has been added to make the model to avoid unnecessary processing of the results which are not in top "n".

Each result item retrieved from the component $\lambda$ is parsed by the parser as shown in (4).

$$P(\lambda) = P\left[\left\{\frac{\Gamma}{\delta}\right\}^{\varepsilon}\right] = \langle \alpha_1, \alpha_2, \alpha_3 .... \alpha_\varepsilon \rangle \quad (4)$$

In (4), $\langle \alpha_1, \alpha_2, \alpha_3 .... \alpha_\varepsilon \rangle$ denotes the resultant items retrieved by $\lambda$. Hence these result items are retrieved from social networks they inherit a particular structure. From this structure the user related information corresponding to the e-mail address provided would be parsed and fetched.

$$\forall \alpha_i, (i = 1...\varepsilon) : \{N, G, P, I\} \quad (5)$$

Each $\alpha_i$ can be parsed to give information corresponding to the e-mail provided by the user. This is represented by a quadruple $\{N, G, P, I\}$.

The four components represent the following.

N = Name

G = Gender

P = Place

I = Image

Each result item retrieved from the component $\pi$ is parsed by the parser as shown in (6).

$$P(\pi) = P\left[\left\{\frac{\Pi}{\delta}\right\}^{\varepsilon}\right] = \langle \beta_1, \beta_2, \beta_3 .... \beta_\varepsilon \rangle \quad (6)$$

In (6), $\langle \beta_1, \beta_2, \beta_3 .... \beta_\varepsilon \rangle$ denotes the resultant items retrieved by $\pi$ . The parsing of the resultant items $\langle \beta_1, \beta_2, \beta_3 .... \beta_\varepsilon \rangle$ is complex than the parsing of $\langle \alpha_1, \alpha_2, \alpha_3 .... \alpha_\varepsilon \rangle$. These web results are used to fetch the blogger profile which would be listed by the visualizer.

B.  *The Algorithm*

The macro-level steps of the algorithm for the proposed model are listed in this section. Each of the steps given can be implemented in different possible methods. Due to this, the micro-level implementation related details are not explored in detail in the algorithm.

**Algorithm WeIPe**

Input   : e-mail id $\delta$

Output : Fetched information corresponding to the e-mail id.

**Begin**

1. Connect to Social Networks
2. Use the Social Networks API to fetch information on given e-mail
3. Parse the results to locate required data
4. Consolidate the information fetched from SN API
5. Connect to the Web resources
6. Fetch using API of traditional search engines
7. Parse results for profile related information
8. Supply the output generated in (4) and (7) to visualizer
9. Render the data gathered using a multi-tab interface.

**End**

## IV. EXPERIMENTATION AND RESULTS

This section would highlight the experimental setup used for the validation of above mentioned model and algorithms. The prototype implementation is done with the software stack including Linux, Apache, MySql and PHP. For client side scripting JavaScript is used. With respect to the hardware, a dual processor system with 3 GHz of speed, 6 GB of RAM is used. The internet connection used in the experimental setup is a 64 Mbps leased line.

A screenshot of the implementation of We.I.Pe model is shown in Figure 2 and Figure 3.

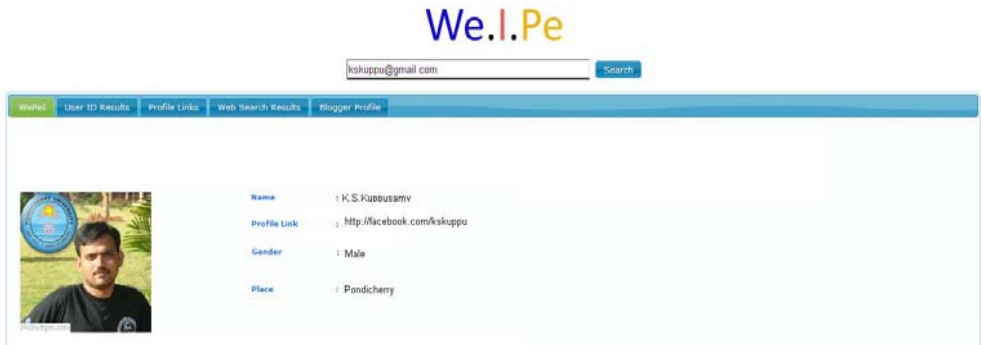

Figure 2.  A Screenshot of We.I.Pe Implementation – Summary Information

Figure 2 depicts the summary information retrieved using social networks and Figure 3 depicts blog profile information.

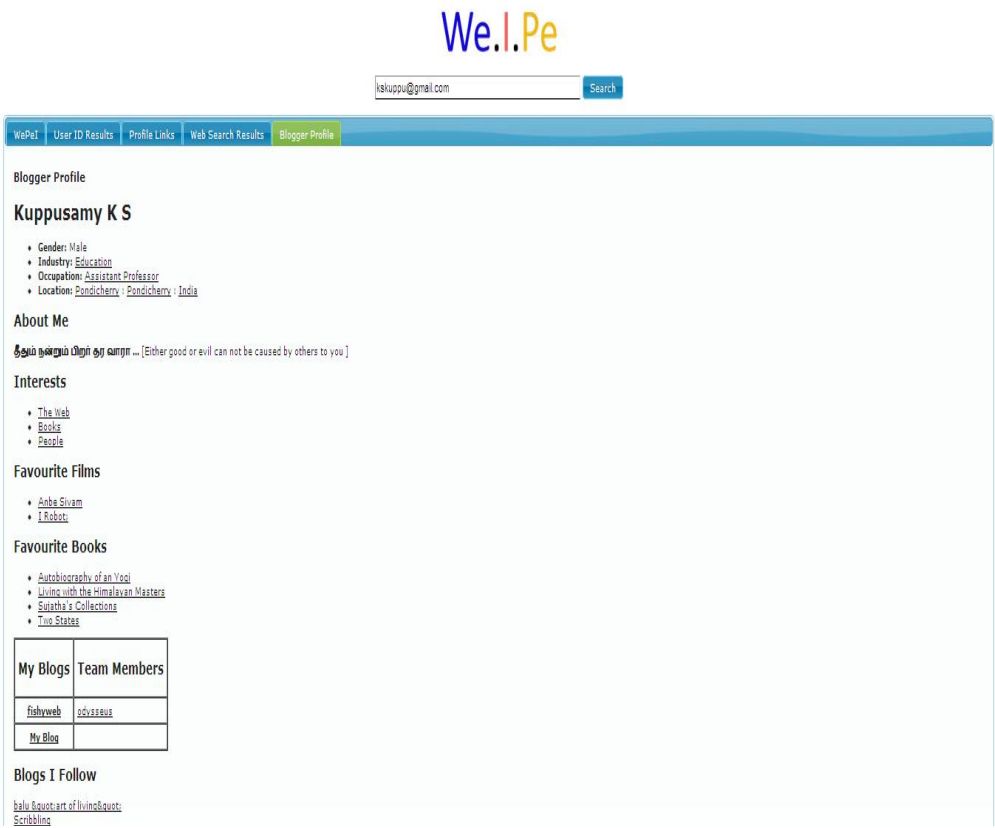

Figure 3.  A Screenshot of We.I.Pe Implementation – Blog Profile Information

The experiments were conducted on the prototype implementation in various sessions. The results for ten different sessions are depicted in Table 1. To compare the results each session is made to have 20 different searches.

TABLE I. PERFORMANCE ANALYSIS

| Session | Count of Successful Searches | |
|---|---|---|
| | *Summary Information* | *Blogger Profile* |
| 1 | 15 | 10 |
| 2 | 17 | 10 |
| 3 | 18 | 13 |
| 4 | 16 | 14 |
| 5 | 15 | 11 |
| 6 | 10 | 8 |
| 7 | 7 | 7 |
| 8 | 17 | 12 |
| 9 | 14 | 13 |
| 10 | 10 | 6 |

The results shown in Table 1 are plotted as chart in Figure 4.

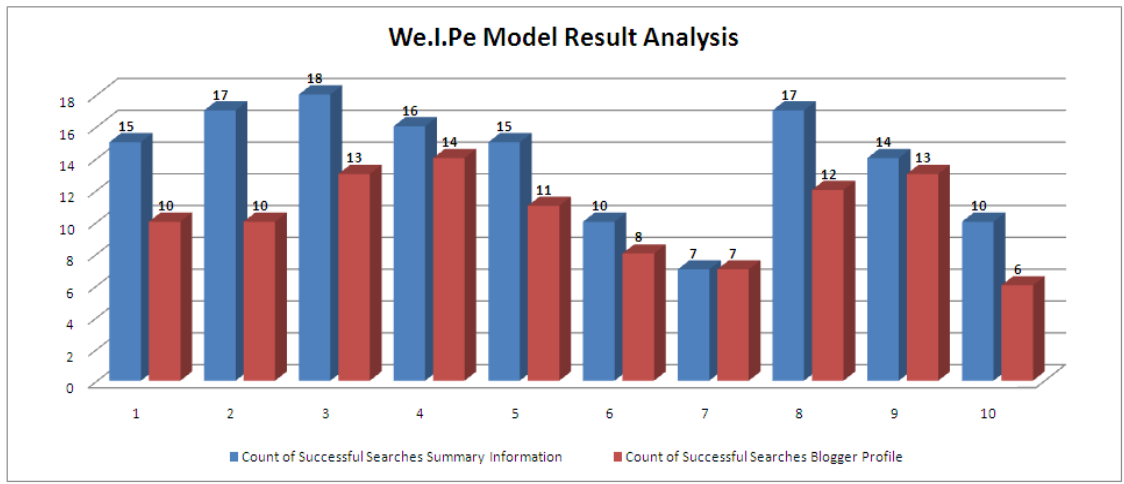

Figure 4. The We.I.Pe model result analysis

From the chart in Figure 4 it can be observed that in the entire search sessions the summary information is successfully retrieved in more instances comparing the blog profile. The reason for the above is that many of the e-mail id don't have an associated blog profile.

## V. CONCLUSIONS AND FUTURE DIRECTIONS

The proposed We.I.Pe model makes the identification of information pertaining to an e-mail id simpler and effective. Hence the information is gathered from various sources including social networks and other web resources the success rate of data is considerably higher. From the experiments conducted it is observed that around 70% of searches returned positive results pertaining to the e-mail id.

The future directions for this research work include incorporation of more details pertaining to the user like work place, subjects of interest etc.

### ACKNOWLEDGMENT

The authors would like to acknowledge the efforts taken by Mr. Suresh, a Post Graduate Student of the department in implementing the We.I.Pe model.

### REFERENCES

[1] R. Guha and A. Garg. Disambiguating people in search. In Stanford University, 2004.
[2] http://people.yahoo.com/ retrieved on 20th March 2011.
[3] http://pipl.com retrieved on 20th March 2011.
[4] http://www.intelius.com/people-search.html retrieved on 20th March 2011/
[5] Artiles, J., Gonzalo, J., Verdejo, F.: A testbed for people searching strategies in the WWW. In SIGIR(2005) 569-570.
[6] K. Balog, L. Azzopardi, and M. de Rijke. Personal name resolution of web people search. In WWW2008 Workshop: NLP Challenges in the Information Explosion Era (NLPIX 2008), 2008.
[7] Chen, Ying, James Martin and Martha Palmer, "Robust Disambiguation of Web-based Personal Names," In Proc. of Second IEEE International Conference on Semantic Computing, 2008.
[8] Popescu, Octavian and Bernardo Magnini, "IRST-BP: Web People Search Using Named Entities," In Proceedings of Semeval 2007, Association for Computational Linguistics, 2007.
[9] Javier Artiles, Enrique Amigo; and Julio Gonzalo. 2009. The role of named entities in web people search. In Proceedings of the 2009 Conference on Empirical Methods in Natural Language Processing: Volume 2 - Volume 2 (EMNLP '09), Vol. 2. Association for Computational Linguistics, Stroudsburg, PA, USA, 534-542.
[10] Vacharasintopchai, Thiti;Nguyen-Huu, Phong, "An Improved People-Search Technique for Directed Social Network Graphs" in Proc. 2nd International Conference on Robotics, Informatics, and Intelligent Technology (RIIT2009).

AUTHORS PROFILE

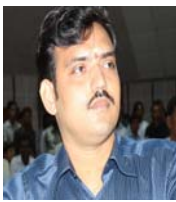

**K.S.Kuppusamy** is an Assistant Professor at Department of Computer Science, School of Engineering and Technology, Pondicherry University, Pondicherry, India. He has obtained his Masters degree in Computer Science and Information Technology from Madurai Kamaraj University. He is currently pursuing his Ph.D in the field of Intelligent Information Management. His research interest includes Web Search Engines, Semantic Web.

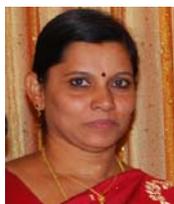

**G. Aghila** is a Professor at Department of Computer Science, School of Engineering and Technology, Pondicherry University, Pondicherry, India. She has got a total of 20 years of teaching experience. She has received her M.E (Computer Science and Engineering) and Ph.D. from Anna University, Chennai, India.

She has published nearly 50 research papers in web crawlers, ontology based information retrieval. She is currently a supervisor guiding 8 Ph.D. scholars. She was in receipt of Schrneiger award. She is an expert in ontology development. Her area of interest include Intelligent Information Management, artificial intelligence, text mining and semantic web technologies.